\documentclass[mathleft
]{an}
\usepackage{graphicx}
\usepackage{times}
\usepackage{graphics,graphicx}
\usepackage{paralist}
\usepackage{amsmath}
\usepackage{amssymb}

\usepackage{natbib}
\begin{document}

\sloppypar

\Pagespan{1}{}
\Yearpublication{2010}%
\Yearsubmission{2010}%
\Month{}%
\Volume{}%
\Issue{}%

\title{The collective X-ray luminosity of HMXB as a SFR indicator}

\author{S. Mineo\inst{1}\fnmsep\thanks{Corresponding author: \email{mineo@mpa-garching.mpg.de}\newline}, M. Gilfanov\inst{1,2}  \and R. Sunyaev\inst{1,2} 
}
\titlerunning{The collective X-ray luminosity of HMXB as a SFR indicator}
\authorrunning{S. Mineo, M. Gilfanov \& R. Sunyaev}
\institute{
Max Planck Institut f\"ur Astrophysik, Karl-Schwarzschild-Str. 1 85741 Garching, Germany
\and 
Space Research Institute of Russian Accademy of Sciences, Profsoyuznaya 84/32, 117997 Moscow, Russia
}

\received{31 Aug 2010}
\accepted{20 Sep 2010}
\publonline{later}


\abstract{%
We study the relation between the X-ray luminosity of compact sources and the SFR of the host galaxy. Our sample includes 38 galaxies for which a uniform set of X-ray, infra-red and ultraviolet data from \textit{Chandra}, Spitzer and GALEX has been collected. Our primary goals are (i) to obtain a more accurate calibration of the $L_{\mathrm{X}}-\mathrm{SFR}$ relation and (ii) to understand the origin of the dispersion in the $L_{\mathrm{X}}-\mathrm{SFR}$ relation observed in previous studies. Preliminary results of this project are reported below.}

\maketitle
\section{Introduction}
The recent star formation rate (SFR) of a galaxy is measured from the light emitted by massive stars: given their short lifetimes, their luminosity is, to the first approximation, proportional to the rate at which they are currently being formed.
The conversions between luminosities at different wavelengths and SFR are usually derived from stellar population synthesis models \citep[see review by][]{Ken98}. Most of the SFR indicators are known to be subject to various uncertainties, one of which is correction for the interstellar extinction. 

Another method of SFR determination is based on the X-ray emission of a galaxy. As X-rays are able to penetrate large columns of gas and dust, such an estimator is expected to be independent of this uncertainty of the more conventional indicators. The X-ray emission of a normal gas-poor galaxy is  dominated by the collective emission of its X-ray binary (XRB) population \citep[e.g.][and references therein]{Fab06}. The latter can be divided into  low- and high-mass X-ray binaries according to the mass of the donor star. The difference in mass results in different evolution time-scales. The characteristic time-scale of luminous high-mass X-ray binaries (HMXB) does not exceed $\sim {\rm few}\times 10^{7}\,\mathrm{yr}$, making them a potentially good tracer of the recent star formation activity in the host galaxy. 

Compact stellar mass X-ray sources more luminous than $\sim 10^{39}$ erg/s are named ULXs \citep[ultra-luminous X-ray sources,][]{Mushotzky04}. They are predominantly located in star-forming galaxies. \cite{Grimm03} showed that ULXs form the bright end extension of the luminosity function of compact sources in star-forming galaxies, the fainter end of which is populated by "normal" HMXBs. Due to relatively shallow slope of the HMXB XLF,  the collective X-ray luminosity of compact sources in a galaxy is dominated by its brightest sources, thus ULXs drive the correlation between X-ray luminosity and the SFR of the host galaxy.

In the past decade several authors have discussed the X-ray emission of compact sources as an estimator of the SFR of the host galaxy \citep[e.g.][]{Grimm03,Ranalli03,Persic07}. Their work has established the general correlation between the X-ray luminosity of the galaxy (and/or of its binary population) and its SFR. However some discrepancy between the scale factors derived by different authors remained. Moreover, a rather large scatter in the $L_{\mathrm{X}}$/SFR ratio has been seen in the individual samples. One of the limiting factors of previous studies was the heterogeneous nature of the X-ray and SFR data. With the large numbers of galaxies available nowadays in \textit{Chandra}, Spitzer and GALEX public data archives it is became possible to construct  large samples of galaxies for which homogeneous sets of multiwavelength measurements are available. This motivated us to revisit the problem of the $L_{\mathrm{X}}-$SFR relation in order to (i) obtain its more accurate calibration and (ii) understand  the origin of the dispersion seen in previous studies.

\section{The sample}
We started to build the sample of star-forming galaxies from the entire "Normal 
Galaxies" section of the \textit{Chandra} Data Archive. It includes over 300 galaxies of all
morphological types. We based our selection on the following four criteria.

\begin{inparaenum}[\itshape i\upshape)]
\item \textit{Hubble type:} we selected only late-type galaxies, i.e. those having numerical index of stage along the Hubble sequence $T>0$;

\item \textit{Specific SFR:} as  star-forming galaxies contain also LMXBs, unrelated to the recent star-formation activity, the contribution of the latter will contaminate the $L_X-$SFR relation. Although separation of the high- and low- mass XRBs in external galaxies is virtually impossible, the LMXBs contribution can be controlled in a statistical way. To this end we use the fact that their population is proportional  to the stellar mass $M_{\star}$ of the host galaxy \citep[][]{Gilfanov04}, whereas HMXBs scale with SFR \citep{Grimm03}, therefore the specific SFR ($\mathrm{SFR}/M_{\star}$) can serve as a measure of their relative contributions to the X-ray luminosity of the galaxy. Based on the scale-factors for HMXB and LMXB populations we estimated that a threshold of $\mathrm{SFR}/M_{\star} >1 \times 10^{-10}$ yr$^{-1}$ would be a reasonable compromise, sufficient for our purpose. In applying this criterium we estimated the $M_*$ and SFR from 2MASS and IRAS catalogues respectively.

\item \textit{Number of compact sources:}  in order to perform the spatial analysis of the detected sources (see Sect. 3), we required that galaxies in our sample have 10 or more compact sources within the isophotal diameter $D25$. After applying this criterium, all  remaining galaxies had X-ray point source detection sensitivity of  $L_{\mathrm{lim}} < 5\times 10^{37}$ erg s$^{-1}$;

\item \textit{Distance:} for the galaxies passed through the above criteria,  a number of redshift independent distances, determined with different methods, were collected from NED. 
Following the results of \cite{Jacoby92}, we chose the most accurate among them.
For our primary sample we chose a distance threshold of 40 Mpc, in order to spatially 
resolve the X-ray compact sources with \textit{Chandra}. This also allows us to identify and separate the central AGN from the rest of detected point sources. The thus defined primary sample includes 29 galaxies, with SFR values ranging from 0.1 to 15 $M_{\odot}$ yr$^{-1}$ in the $0.5-8\,\mathrm{keV}$ band.

{\em High-SFR sample.}
In order to explore the $L_{\mathrm{X}}-\mathrm{SFR}$ relation at  higher SFR, we relaxed the distance constraint and constructed the secondary high-SFR sample of more distant galaxies. A total number of 8 galaxies were found beyond the distance of $D>40$ Mpc with the SFR in the $13-412$ $M_{\odot}$ yr$^{-1}$ range. They were treated as spatially unresolved in our analysis (see section 4 for details and caveats). Five of them are classified as ultra-luminous infrared galaxies (ULIRGs) and two as luminous infrared galaxies (LIRGs). Although they were contained in the normal galaxies section of the \textit{Chandra} archive, we verified that the contribution of AGN to the bolometric luminosity in these galaxies is $< 30 \%$ \citep[][]{Nardini09}, to ensure that the IR-based SFR measurement is not compromised by the latter. The AGN contribution in the X-ray band was constrained directly from the \textit{Chandra} images (these galaxies are partly resolved by \textit{Chandra}, the smallest one having angular size of $D25\sim24$ arcsec). Finally, we included 8 normal galaxies from the Chandra Deep Field North \citep[][]{Bra01}. The K-corrected X-ray and radio luminosities for these galaxies were taken from \cite{Gilf04}. Thus, our high-SFR sample included 16 galaxies.
\end{inparaenum}

\begin{figure*}
 \centering
\begin{minipage}{155mm}
\hbox
{
\includegraphics[width=75mm]{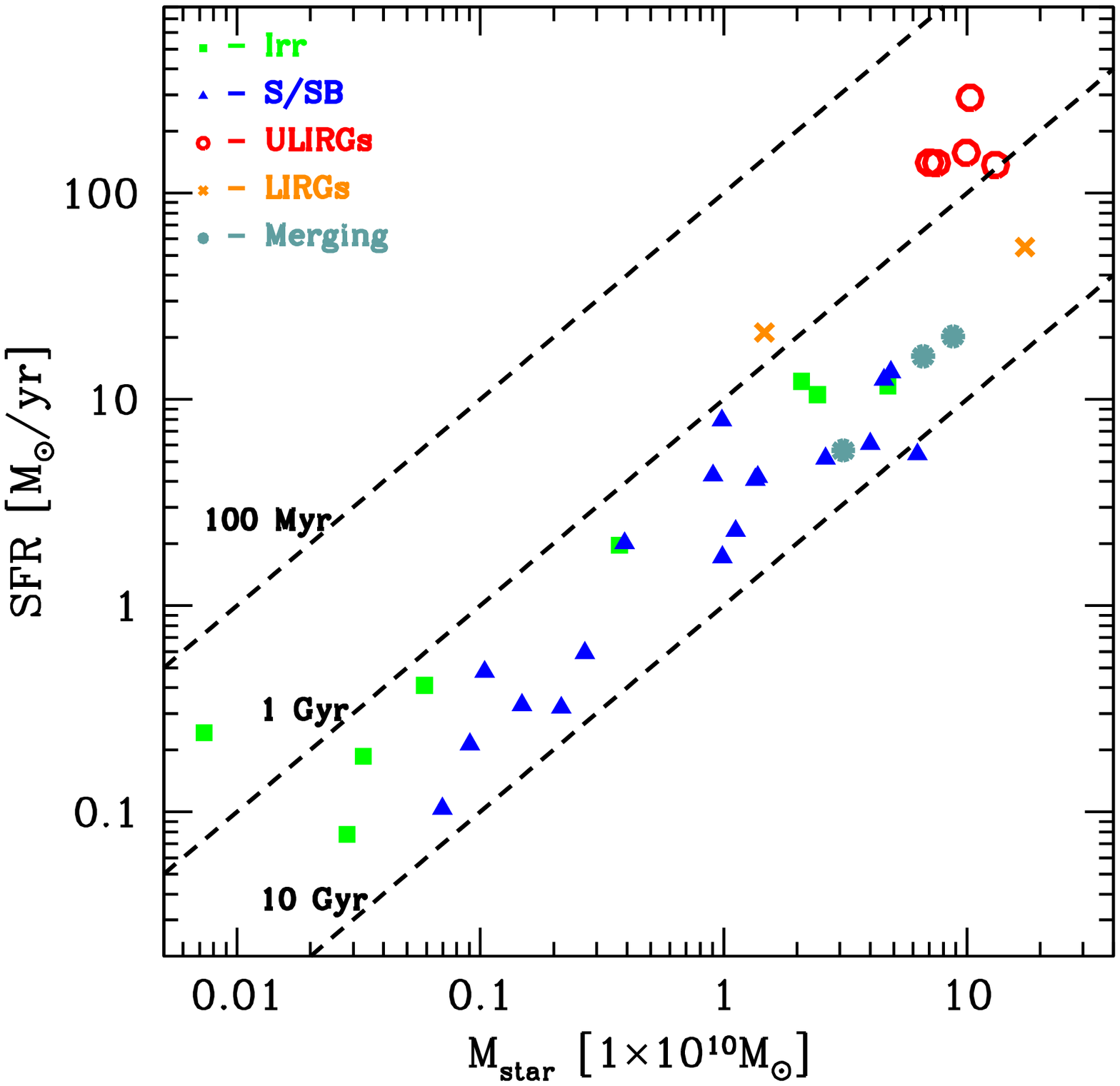}
\includegraphics[width=75mm]{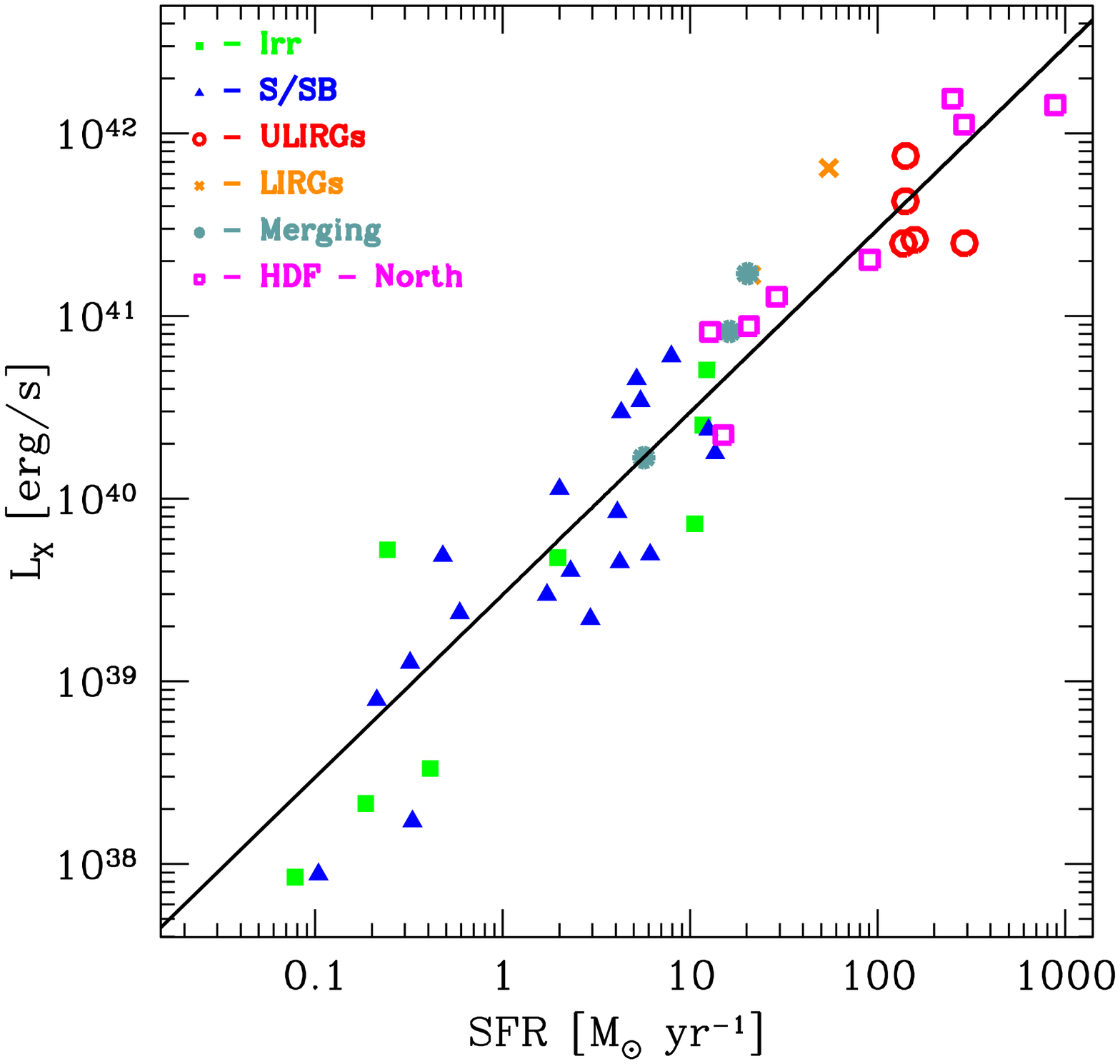}
}
\caption{\footnotesize \emph{Left:} The $\mathrm{SFR}-M_{\star}$ plane. Different types of galaxies are plotted with different symbols, together with the lines of constant stellar-mass-to-SFR ratio (dashed lines). \emph{Right:} The $L_{\mathrm{X}}-\mathrm{SFR}$ relation. The solid line shows the linear relation given by the equation in Sec.5.} 
\label{fig:lx_sfr}
\end{minipage}
\end{figure*}

\section{Spatial analysis}
In order to identify regions of galaxies where the HMXBs dominate the compact X-ray source population, we analyzed the spatial distribution of the detected compact sources in resolved galaxies of the primary sample.

To minimize the contribution of CXB sources we constructed cumulative and differential radial profiles of the detected X-ray point sources.
These were compared  with the surface density of CXB sources  predicted from $\log N-\log S$ of \cite{Georg08}. Based on this analysis we chose regions where the predicted number of CXB sources did not exceed $30\%$ of the total number of detected sources.

Although only star-forming galaxies with $\mathrm{SFR}/M_{\star}\geq 1 \times 10^{-10}$ yr$^{-1}$ have been selected, we further minimized the contribution of LMXBs to the total X-ray luminosity by  
locating the bulge regions of spiral galaxies and excluding them from X-ray and multiwavelength analysis.

\section{Multiwavelength analysis and SFR estimation}
The X-ray luminosity for galaxies in our primary sample was measured as the sum of luminosities of point sources detected above the corresponding sensitivity limit. As \textit{Chandra} observations of different galaxies achieved different sensitivity, these luminosities need to be transformed to the same threshold.  The luminosity  of "missing" unresolved sources was computed  by integrating the luminosity function which differential slope was assumed to be $1.6$ \citep[][]{Grimm03}. The normalization was determined for each galaxy  individually, from the number of sources detected above its sensitivity limit.
We chose  $L_{\mathrm{X}} = 10^{36}$ erg/s as the common luminosity threshold, i.e. the luminosities presented below are an estimate of the combined luminosity of compact sources brighter that this threshold. The count-to-erg conversion was done assuming a power law spectrum with slope of $2.0$ and absorption of $n_{H}=3\times 10^{21}$ cm$^{-2}$. 

The galaxies from the secondary high-SFR sample were treated as unresolved objects and their luminosity was measured as total luminosity inside the $D25$ ellipse.  The counts-to-ergs conversion in this case was determined based on the modeling of their X-ray spectra.

For the purpose of studying the $L_X-$SFR relation for X-ray binaries we have  subtracted from the total luminosity of point sources in the primary sample the contribution of the CXB sources. The latter was estimated statistically, based on the CXB $\log N-\log S$ of \cite{Georg08}. The resulting correction does not exceed $\sim 35\%$. This correction is insignificant for unresolved galaxies from the high-SFR sample.

We used data from Spitzer, GALEX and 2MASS archives to obtain accurate and homogeneous estimation of SFR and stellar mass.
For resolved galaxies we measured fluxes of the FIR ($70\,\mu$m and $24\,\mu$m) , NIR ($K_{\mathrm{S}}$ band) and UV (FUV and NUV) within the regions selected as described in Sect. 3. For unresolved galaxies from the high-SFR sample we measured fluxes within the $D25$ ellipse.  

We computed  the SFR of each galaxy  combining  IR- and UV- based estimates \citep[see][]{Hira03,Igl06}. This method accounts for the light emitted by the dust heated by young stars as well as the UV photons escaping the dust clouds. It has the advantage of being free of the model dependent  attenuation correction of the UV emission. We  also included a correction  for  the IR emission from  old stellar population: 
\begin{equation}
\label{eq:sfr_tot}
 \mathrm{SFR} = \mathrm{SFR_{UV}^{0}}+(1-\eta) \mathrm{SFR_{IR}}
\end{equation}
where $\mathrm{SFR_{UV}^{0}}$ and $\mathrm{SFR_{IR}}$ are computed using standard calibrations for the Salpeter 
IMF from 0.1 to 100 $M_{\odot}$  \citep[][]{Ken98p}:
\begin{equation}
\label{eq:sfr_ir}
 \mathrm{SFR_{IR}} (M_{\odot}\mathrm{ yr}^{-1}) = 4.6 \times 10^{-44} L_{\mathrm{IR}} (\mathrm{erg}\,\mathrm{s}^{-1})
\end{equation}
\begin{equation}
\label{eq:sfr_nuv_obs}
\mathrm{SFR_{UV}^{0}} (M_{\odot}\mathrm{ yr}^{-1}) = 1.2 \times 10^{-43} L_{\mathrm{NUV,obs}} (\mathrm{erg}\,\mathrm{s}^{-1})
\end{equation}
$L_{\mathrm{ NUV,obs}}$ is the luminosity at $2312\,\AA$ un-corrected for dust 
attenuation, $L_{\mathrm{IR}}$ is the $8-1000\,\mu$m luminosity, $\eta$ is the cirrus correction, i.e. the fractional contribution  of the old stellar population to $L_{\mathrm{IR}}$. We used standard values of  $\eta=0$ for starbursts and $\eta=0.4$ for disk galaxies.

The SFR for HDFN galaxies were computed based on their 1.4 GHz luminosities and calibration of \cite{Bell03}:
\begin{equation}
\label{eq:sfr_radio}
 \mathrm{SFR} (M_{\odot}\mathrm{ yr}^{-1}) = 5.52\times 10^{-29}L_{1.4\,\mathrm{GHz}}(\mathrm{erg}\,\mathrm{s}^{-1})
\end{equation}

\section{\boldmath $L_{\mathrm{X}}-\mathrm{SFR}$ relation}

The obtained  $L_{\mathrm{X}}-\mathrm{SFR}$ relation is shown in the right-hand panel of Fig. \ref{fig:lx_sfr}. An important caveat is that we have plotted galaxies from both resolved and high-SFR samples in the same plot, ignoring the fact that X-ray luminosity of the latter includes contribution of unresolved emission (Sect. 4), which potentially may be significant in star-forming galaxies. This issue will be fully addressed in the future publication. Our preliminary estimates show that contribution of unresolved emission does not change the X-ray luminosity of star-forming galaxies by a large factor. This is consistent with the fact that galaxies from the unresolved sample appear to lie on the extension of the relation for resolved galaxies.

It has been shown that  $L_{\mathrm{X}}-$SFR relation for discrete sources is subject to effects of statistic of small numbers which tend to steepen the apparent slope of its low-SFR part \citep[][]{statistical04}. Therefore a fit to the relation in the entire SFR range is likely to produce biased results, depending on the distribution of the galaxies of the sample across SFR. We defer detailed consideration of this effect to  the forthcoming publication and  plot for the reference in the Fig.\ref{fig:lx_sfr} a linear relation 
$L_\mathrm{X}(\mathrm{erg}\,\mathrm{s}^{-1}) = 3\times 10^{39}\,\mathrm{SFR}(M_{\odot}\,\mathrm{yr}^{-1})$

\section{\boldmath Dispersion around the $L_{\mathrm{X}}-\mathrm{SFR}$ relation}
\begin{figure*}
\centering
 \begin{minipage}{140mm}
\hbox
{
\includegraphics[width=45mm]{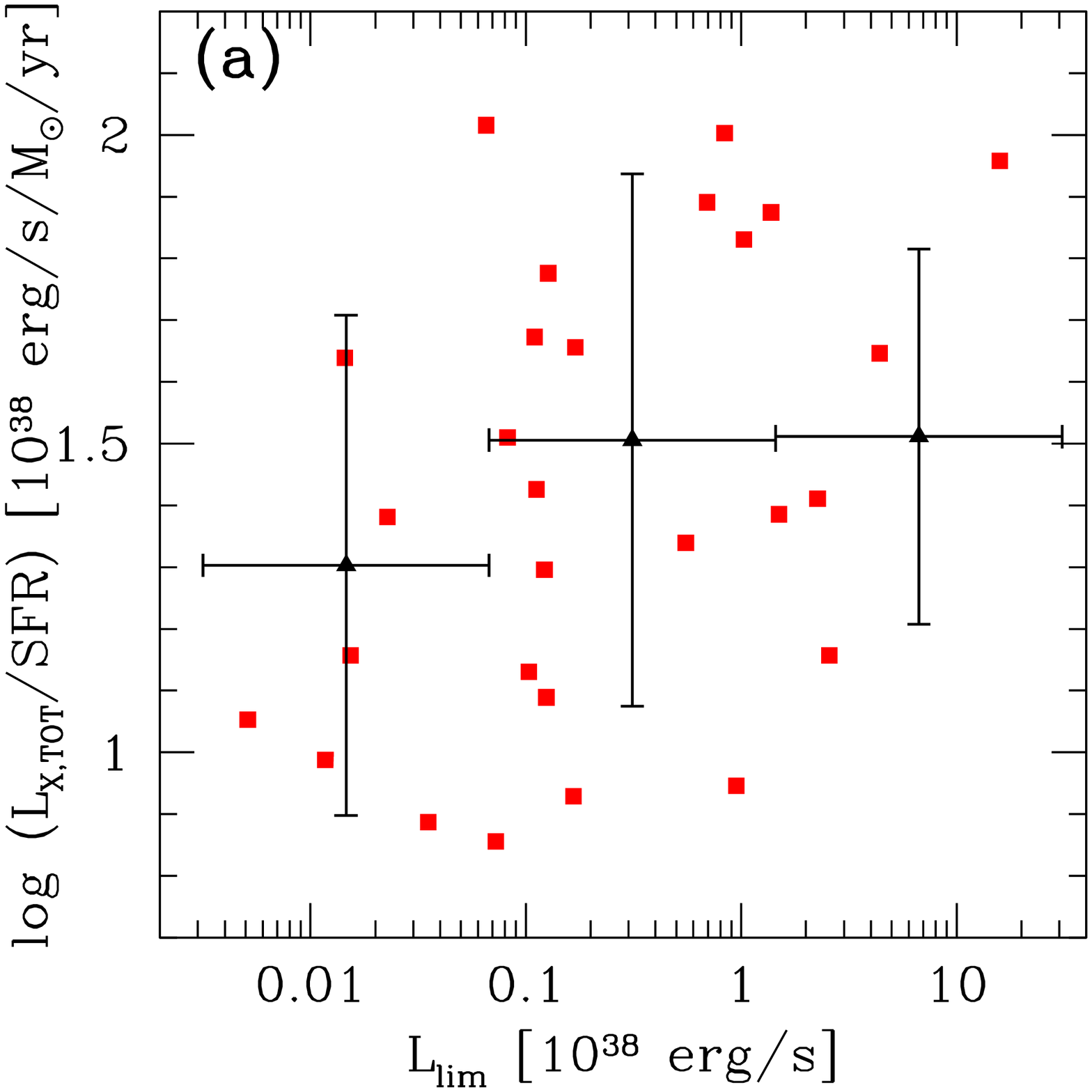}
\includegraphics[width=45mm]{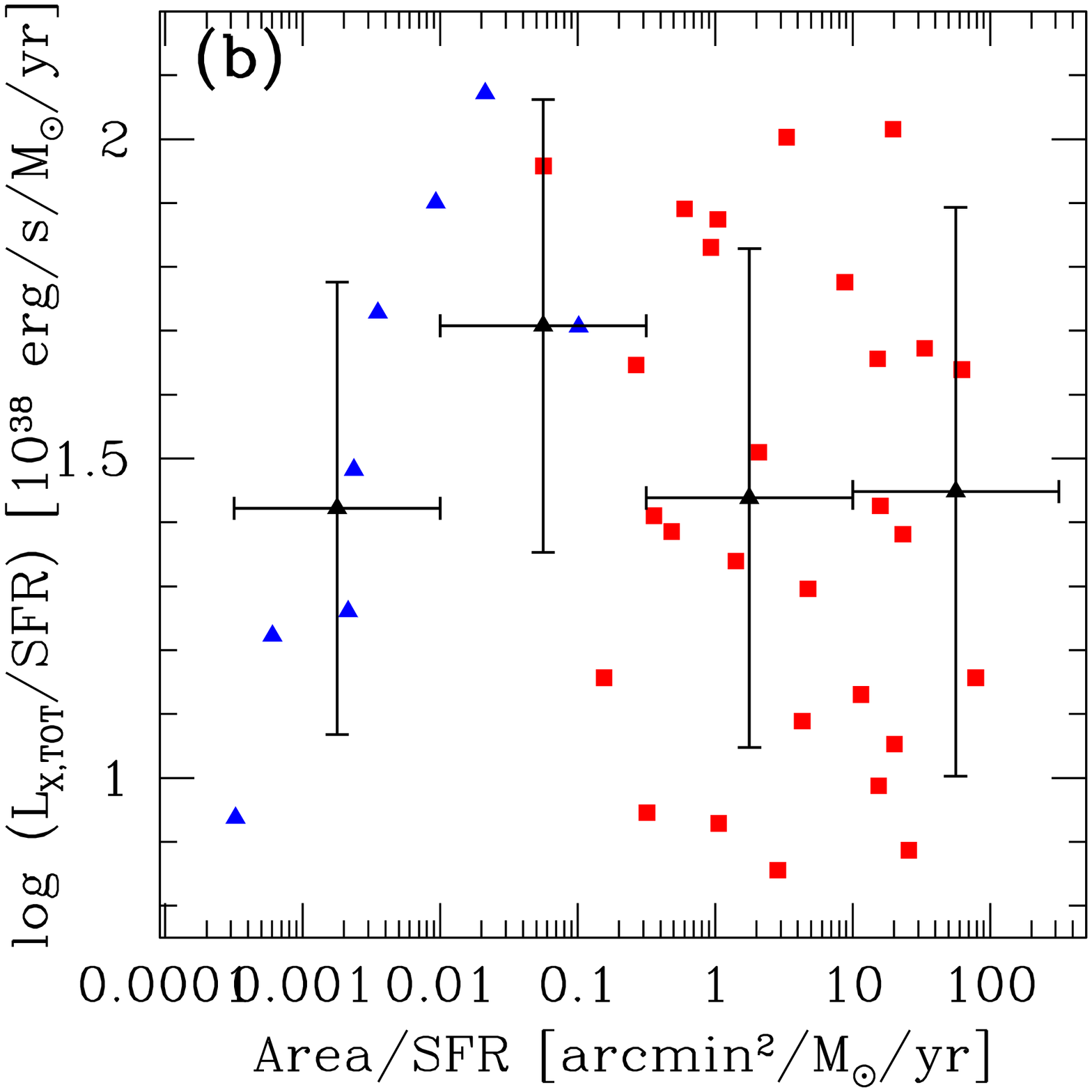}
\includegraphics[width=45mm]{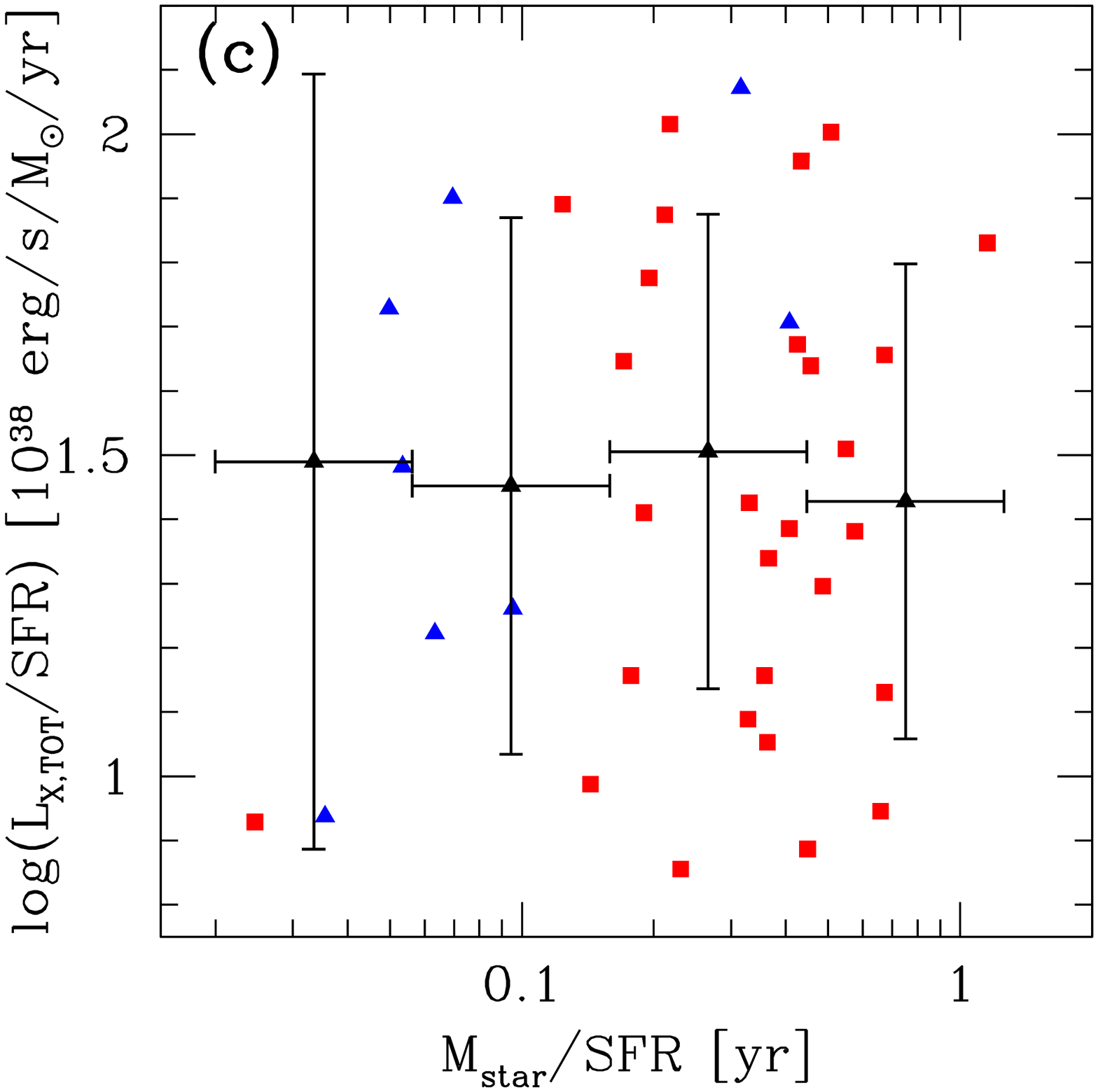}
}
\centering
{
\includegraphics[width=45mm]{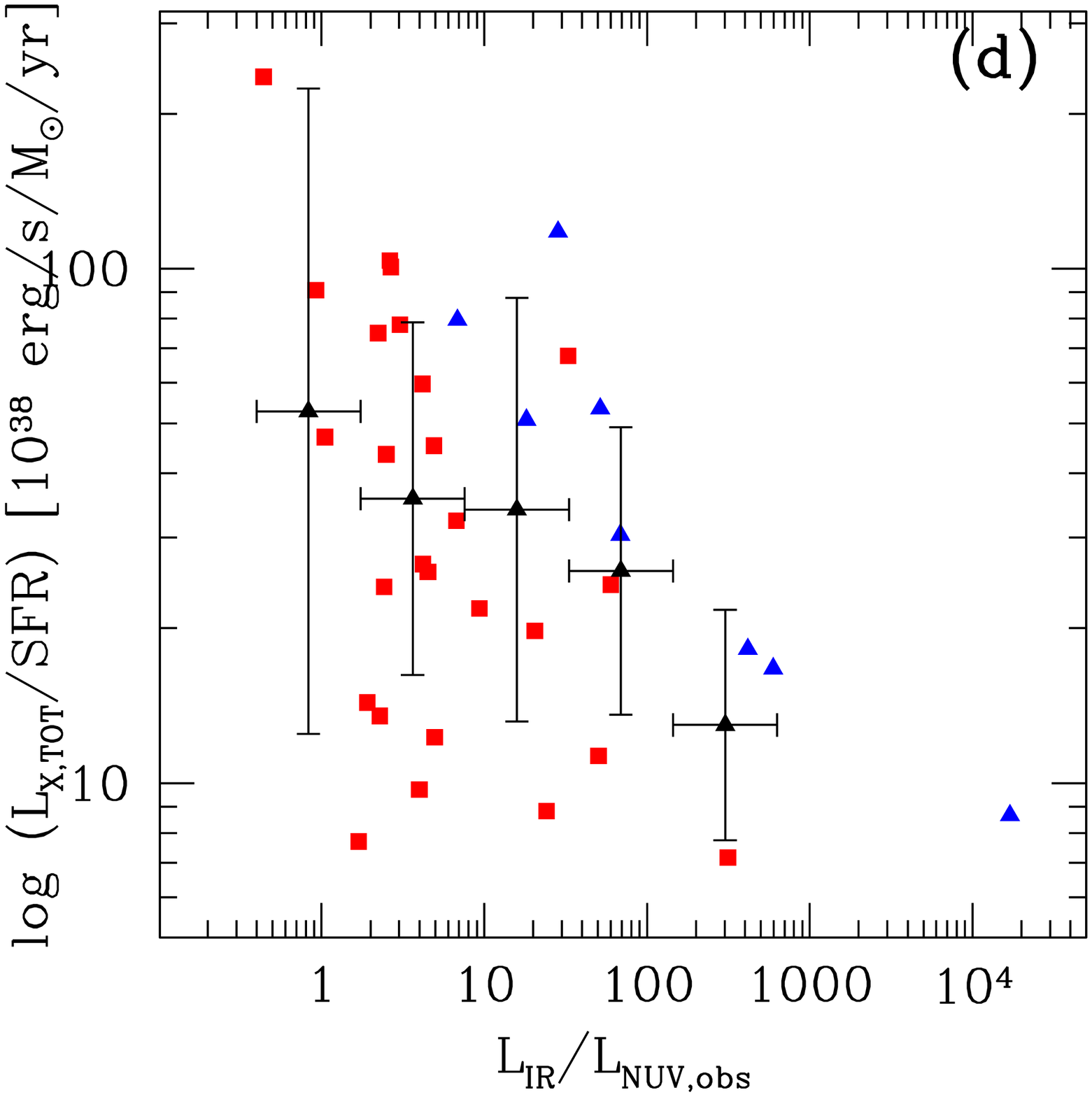}
\includegraphics[width=45mm]{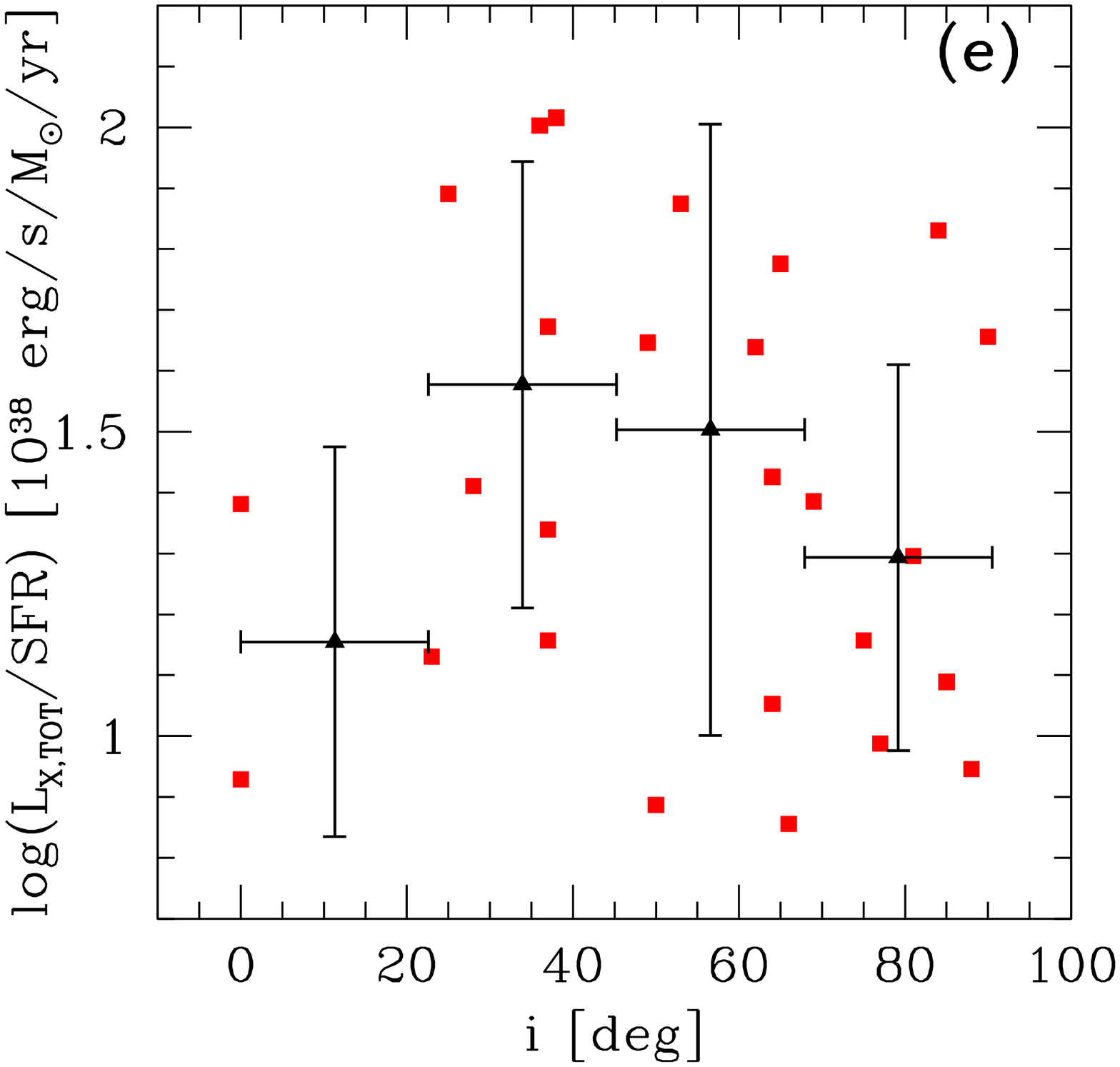}
}
\caption{\footnotesize Search for correlation of $L_X/$SFR ratio with various potential contaminating factors: variations in sensitivity limit of X-ray data (a);  contribution of CXB sources (b) and LMXBs (c); dust attenuation (d) and  inclination of the galaxy (e). Resolved galaxies are shown by squares, unresolved -- with triangles. See Sect. 6 for details.}
\label{fig:dispersion_study}
\end{minipage}
\end{figure*}

Despite of the special effort to  create a sample with a homogeneous set of multi-wavelength measurements and to minimize contamination by CXB sources and LMXBs, the resulting $L_{\mathrm{X}}-$SFR relation bears a rather large dispersion of $\sim 0.4$ dex rms.
In order to investigate the origin of this dispersion we searched for correlation of the $L_{\mathrm{X}}/\mathrm{SFR}$ ratio with different quantities characterizing importance of various potential contaminating factors. These included the sensitivity limit of \textit{Chandra} data $L_{\mathrm{lim}}$, contribution of CXB and LMXB sources characterized by the ratio of the studied area and enclosed stellar mass to the SFR, dust attenuation characterized by the $L_{\mathrm{IR}}/L_{\mathrm{UV}}$ ratio and inclination of the galaxy $i$.  The results are shown in Fig. \ref{fig:dispersion_study}. In these plots, the symbols without error bars represent individual galaxies, and symbols with error bars show binned values. The vertical error bars represent the bin rms. 
We searched for statistically significant correlations in these data using the Spearman's rank correlation test and found none, with the possible exception of the dust attenuation where the associated probability was on the $\sim 0.4-6\%$ level, depending on the samples combination. As it is obvious from the plot, the correlation is such that galaxies with stronger dust attenuation have smaller $L_{\mathrm{X}}/$SFR ratio. This is to be expected. We note however, that in order to suppress the 0.5--8 keV luminosity by a factor of 10, a column density of $\sim5\times10^{23}$ cm$^{-2}$ is needed.

\section{Conclusions}
We presented preliminary results of the study of the relation between SFR and X-ray luminosity of galaxies based on \emph{Chandra}, Spitzer, GALEX and 2MASS observations of nearby star-forming galaxies. We found an $L_{\mathrm{X}}-\mathrm{SFR}$ relation that is in a general agreement with results of earlier studies. Despite our attempt to minimize the contribution of various contaminating factors, we still see the relatively large dispersion in the $L_{\mathrm{X}}-\mathrm{SFR}$ relation, of the order of $\sim 0.4$ dex rms. We investigated the possible role of various effects and concluded that the observed dispersion is unlikely to be caused by any of the obvious contaminating factors and may have a physical origin. Plausible candidates are differences in the recent star-formation history, metallicity, attenuation by the dust and gas etc.

\acknowledgements
\tiny{This research made use of \textit{Chandra} archival data provided by the \textit{Chandra} X-ray Center in the application package CIAO. This publication makes use of data products from Two Micron All Sky Survey, which is a joint project of the University of Massachusetts and the Infrared Processing and Analysis Center/California Institute of Technology, funded by the NASA and the National Science Foundation. The \textit{Spitzer Space Telescope} is operated by the Jet Propulsion Laboratory, California Institute of Technology, under contract with the NASA. \textit{GALEX} is a NASA Small Explorer. We gratefully acknowledge NASAÕs support for construction, operation and science analysis for the \textit{GALEX} mission, developed in cooperation with the Centre National dÕEtudes Spatiales of France and the Korean Ministry of Science and Technology. This research has made use of the NASA/IPAC Extragalactic Database (NED) which is operated by the Jet Propulsion Laboratory, Caltech, under contract with the NASA.}

\end{document}